\documentstyle[12pt]{article}
\newcommand{\be}{\begin{equation}}
\newcommand{\bef}{\begin{figure}}
\newcommand{\eef}{\end{figure}}

\newcommand{\ee}{\end{equation}}
\def\spose#1{\hbox to 0pt{#1\hss}}
\def\ltapprox{\mathrel{\spose{\lower
3pt\hbox{$\mathchar"218$}}
 \raise 2.0pt\hbox{$\mathchar"13C$}}}
\def\gtapprox{\mathrel{\spose{\lower
3pt\hbox{$\mathchar"218$}}
\raise 2.0pt\hbox{$\mathchar"13E$}}}
\def\inapprox{\mathrel{\spose{\lower
3pt\hbox{$\mathchar"218$}}
 \raise 2.0pt\hbox{$\mathchar"232$}}}

\begin{document}

\title{\bf Conceptual Problems of Fractal Cosmology}

\author{Yurij V. Baryshev \\ \\
\small Astronomical  Institute of the Saint-Petersburg University, \\
\small 198904, St.-Petersburg, Russia.  E-mail:  yuba@astro.spbu.ru}

\date{}

\maketitle


\begin{abstract}

This report continues recent Peebles-Turner debate 
"Is cosmology solved?" and considers the first results for Sandage's 
program for "Practical cosmology".
A review of conceptual problems of modern 
cosmological models is given, among them: 
the nature of the space expansion;
recession velocities of distant galaxies more than velocity of light; 
cosmological Friedmann force; continuous creation of gravitating mass 
in Friedmann's equation; 
cosmological pressure is not able to produce a work; 
cosmological gravitational frequency shift; Friedmann-Holtsmark paradox; 
the problem of the cosmological constant; 
Einstein's and Mandelbrot's Cosmological Principles;
fractality of observed galaxy distribution; 
Sandage's 21st problem: Hubble - de Vaucouleurs paradox;
quantum nature of  gravity force.

\end{abstract}

\section{Is cosmology solved?}

A debate under the title "Is cosmology solved?"
was held recently at the Smithsonian
National Museum of National History.  James Peebles
and Michael Turner presented two different views on the problem.
According to Turner(1999) cosmology is solved just in 1998 by the
theory of inflation and cold dark matter. While according to
Peebles(1999) "many commonly discussed elements of cosmology still
are on dangerous ground". Recent discoveries of dominating
contribution of the cosmological constant into the dynamics of the
expansion and fractality of large scale galaxy distribution have
demonstrated how modern powerful observations can change
dramatically common view on cosmological physics.
As Lawrence Krauss said: "One thing is already certain. The standard
cosmology of the 1980s, postulating a flat universe dominated
by matter, is dead." (Krauss,1999). 

Five years ago "23 astronomical problems for the next three decades"
were formulated by Allan Sandage at the conference on Key Problems
in Astronomy and Astrophysics (Sandage,1995). 
The problems N.15 - N.23 relate to "Practical cosmology" and  
recent observations shed light to some of them.
It is clear now that these problems have roots in the foundations of
cosmological models and this is why it is the right
time for an analysis of the basis of contemporary cosmology.

This report is devoted to a continuation of the mentioned above
debate and especially relates to conceptual aspects of cosmological models,
which are sharpened by recent observations and
have been only little discussed previously.

\section{Building blocks of cosmological models}

Any cosmological model contains several
fundamental hypotheses which determine the interpretation
of observable phenomena.
A classification of possible relativistic cosmologies
in accordance with basic initial assumptions  has been discussed
discussed by Baryshev et al.(1994).
Modern cosmological theory includes in particular
as fundamental building blocks
the theory of gravitational interaction, global matter distribution,
origin of cosmic microwave background radiation, mechanism of
cosmological redshift, evolution and the arrow of time.

The most important elements of any cosmological model  are
the {\em cosmological principle} and a {\em relativistic gravity theory},
tying the main conceptual problems of cosmology closely
with recent studies of large scale matter distribution and
investigation of physics of gravitational interaction.

Modern astrophysical observations give the empirical foundation of
cosmological models. The main task of observational cosmology
is to compare predictions of cosmological theories with real
data and to select viable models. During the last decade observations
are developing exponentially and this opens new horizons for
cosmological theory.

Below we give an analysis of contemporary state of
modern cosmology with a special emphasis of conceptual aspects
of cosmological models.

\section{The Standard Model}

The Friedmann-Lemaitre-Robertson-Walker (FLRW) cosmological model
is currently accepted as the Standard Model (SM) for all
interpretations of observed astrophysical data.
For this conference it is interesting to note that
Alexander Friedmann found his famous solution in 1922-1924 working
(partly) here in St.Petersburg University, and at the same time
George Gamow was a student of our university
(together with other brilliant students such as Lev Landau, Dmitrij
Ivanenko, Vladimir Fok, Viktor Ambartzsumyan).

\subsection{Einstein`s Cosmological Principle}

The first basic element  of the SM is  Einstein's
Cosmological Principle. The Cosmological Principle, in fact, is the
hypothesis that the universe is spatially homogeneous and isotropic
on "large scales" (see e.g.  Weinberg 1972; Peebles 1993; Peacock
1999). Homogeneity of the matter distribution plays a central role in the
expanding universe model, because homogeneity implies that the
recession velocity is proportional to distance.  This means that the
linear velocity-distance relation $V=Hr$, identified with the
observed Hubble law, is valid at scales where matter distribution can
be considered on average uniform.  Hence the words "large scales"
have exact meaning in FLRW cosmology as the scales where  linear
velocity-redshift relation starts to exist.

The homogeneity and the isotropy of the matter distribution in space
mean that starting from  scale $r_{hom}$ for all scales
$r>r_{hom}$ we have

\be
\label{rhohom}
\varrho(\vec{r},t) = \varrho(t)
\ee

\be
\label{preshom}
p(\vec{r},t)=p(t)
\ee

It has been extensively discussed whether the homogeneity of
the Universe is to be expected from general physical arguments.
However within the SM one cannot account for the homogeneity
and  this means that homogeneity
must be accepted as a phenomena to be explained by some future deeper
theory.

\subsection{General Relativity}

The second fundamental element of the SM  is the
General Relativity (GR), which is a geometrical gravity theory
(as alternative to the quantum field approach, see e.g. Feynman,1971;
Baryshev,1996).
GR was successfully tested in the weak gravity condition
of the Solar System and binary neutron stars.
It is assumed that GR can be applied to the Universe as a whole.

According to GR gravity is described by a metric tensor
$\:g^{ik}$ of a Riemannian space. The "field" equations in GR
(Einstein-Hilbert equations) have the form:

\be
\label{eheq}
\Re^{ik} - \frac{1}{2}\,g^{ik}\,\Re =
\frac{8\,\pi\,G}{c^4}\,\,T^{ik}_{(m)} + g^{ik}\Lambda
\ee

where $\:\Re^{ik}$ is the Ricci tensor,
$\:T^{ik}_{(m)}$ is the energy-momentum tensor
(hereafter EMT) for the all kinds of matter,
and $\:\Lambda$ is the famous cosmological constant,
which does not depend on time and space coordinates.
Note that gravity in GR is not a
matter, so  $\:T^{ik}_{m}$ does not contain EMT of gravity field.
Solutions of the Eq.\ref{eheq} for unbounded homogeneous matter
distribution (Eqs.\ref{rhohom},\ref{preshom}) are the basis of
FLRW cosmological model.

\subsection{Space expansion paradigm}

An important consequence of homogeneity and isotropy is that
the line element may be presented in the Robertson-Walker form:

\be
\label{rw1}
ds^{2} = c^{2}dt^{2} - S(t)^{2} d\chi^{2} - S(t)^{2} I_k(\chi)^{2}
(d\theta^{2}+\sin^{2}\theta d\phi^{2})
\ee
where $\:\chi,\theta,\phi$ are the "spherical"
comoving space coordinates, $t$ is synchronous time coordinate,
$\:I_k(\chi) = \sin(\chi),\chi,\sinh(\chi)$
corresponding to curvature constant
values $\:k=+1,0,-1$ respectively and $\:S(t)$ is the scale factor.

The {\em expanding space paradigm} is that the
proper metric distance $r$ of a  body with fixed comoving
coordinate  $\:\chi$ from the observer is:

\be
\label{dist1}
r = S(t) \cdot \chi
\ee
and increases with time $\:t$ as the scale factor $\:S(t)$.
Note that physical dimension of metric distance
$\:[r]=cm$, hence if $\:[S]=cm$ then $\:\chi$ is
the dimensionless comoving coordinate distance.
In fact $\chi$ is the spherical angle and $S(t)$ is the radius
of the sphere (or pseudosphere) in the embedding 4-dimensional
Euclidean space. Hence $r$  is "internal" proper distance on the
3-dimensional hypersurface of the embedding space.
In other words  $r$  and $\chi$ are Euler and Lagrangian comoving
distances correspondingly.

Use is often made also of "cylindrical" comoving  space coordinates 
$\mu,\theta,\phi$ , for which the interval is

\be
\label{rw2}
 ds^{2} = c^{2}dt^{2} - S(t)^{2}
\frac{d\mu^{2}}{1-k\mu^2} - S(t)^{2} \mu^{2}
(d\theta^{2}+\sin^{2}\theta d\phi^{2})
\ee

In this case the metric distance $l$

\be
\label{dist2}
l = S(t) \cdot \mu
\ee
is the "external"
distance from z-axis in embedding Euclidean 4-dimensional space.
It is thus important to use different designations for the different
distance interval defined by  Eq.\ref{rw1} and Eq.\ref{rw2}
(see e.g. Peacock,1999,p.70).

The relation between these two
metrical distances is

\be
\label{mdrel}
r = S(t) I^{-1}_k(l/S)
\ee
were $I^{-1}_k$ is the inverse function for $I_k$.

\subsection{Cosmological redshift}

The expansion of space
induces the wave stretching of the traveling photons via 
Lemaitre's equation, i.e.:

\be
\label{lem}
 (1+z) =
 \frac{\lambda_{0}}{\lambda_{1}} = \frac{S_{0}}{S_{1}}
\ee
where $z$ is cosmological redshift,
$\:\lambda_{1}$ and  $\:\lambda_{0}$ are the wavelengths at the
emission and reception, respectively
and $\:S_{1}$ and $\:S_{0}$ the corresponding values of the scale
factor.  Equation (\ref{lem}) is usually obtained from the radial
null-geodesics ($\:ds=0$, $\:d\theta=0$, $\:d\phi=0$) of the RW line
element.

According to the expanding space paradigm, the cosmological redshift
is not the familiar Doppler effect but is a new physical phenomenon
(see discussion in Harrison,1993; 1995). This is clear by
comparison between relativistic Doppler and cosmological FLRW
velocity-redshift relation.

\subsection{Friedmann's equation}

The behavior of the scale factor
 with time $\:S(t)$ is governed by Einstein`s equations
(Eq.\ref{eheq}) which can be written in the form:

\begin{eqnarray}
\Re^k_i - \frac{1}{2}\,\delta^k_i\,\Re = \frac{8\,\pi\,G}{c^4}\,\,T^k_i
\label{eheq2}
\end{eqnarray}

where the total EMT is given by

\begin{equation}
T^k_i = T_{(m)\,i}^k + T_{(r)\,i}^k + T_{(v)_i}^k
\label{emt1}
\end{equation}
Here indexes $m,r,v$ denote matter, radiation and vacuum respectively.
In comoving coordinates the total EMT has the form:

\begin{equation}
T^k_i = diag \left(\varrho c^2, -p, -p, -p \right)
\label{emt}
\end{equation}
where $\varrho=\varrho_m +\varrho_r + \varrho_v$ is the total density
and $p= p_m + p_r + p_v$ is the total pressure. For radiation
$p_r= \frac{1}{3}\varrho_r c^2$
and for vacuum $p_v = -\varrho_v c^2$.

In the case of homogeneity,  Einstein's equations are directly
reduced to  Friedmann's equation, which may be presented in
the following form:

\be
\label{freq1}
\frac{d^{2}S}{dt^{2}}=
- \frac{4 \pi G}{3} S \left( \varrho+ \frac{3p}{c^{2}}\right)
\ee
From the Bianchi identity it follows the continuity equation

\be
\label{cont}
\dot{\varrho} = -3\left(\varrho + \frac{p}{c^2}\right)
\frac{\dot{S}}{S}
\ee
which must be added to  Eq.\ref{freq1}.
Because  Lagrangian comoving coordinates do not depend on time,
one may rewrite Eq.\ref{freq1} using Eq.\ref{dist1} as

\be
\label{freq2}
\frac{d^{2}r}{dt^{2}}= -\frac{GM_g(r)}{r^2}
\ee
where the gravitating mass $M_g(r)$ is given by

\be
\label{mgtot}
M_g = M_m + M_r + M_v
\ee
and contributions from matter, radiation and vacuum are

\be
\label{mgm}
M_m(r) =
\frac{4\pi}{3} \left( \varrho_m + \frac{3p_m}{c^{2}}\right)r^3
\ee
\be
\label{mgr}
M_r(r) = \frac{4 \pi }{3} 2 \varrho_r r^3
\ee
\be
\label{mgv}
M_v(r) = - \frac{4 \pi}{3} 2\varrho_v r^3
\ee

Solving the Friedmann's equation (Eq.\ref{freq2})
one finds the dependence on time for the metric distance
$\:r(t)$ or the scale factor $S(t)$.

\subsection{Cosmological parameters}

The FLRW model has two generally used parameters.
The Hubble parameter $ H = \dot{S}/S$ and the
deceleration parameter $ q = - \ddot{S}S/\dot{S}^2 $
which for the present time
$\:t_{0}$ are $\:H(t_{0})=H_{0}$ and $\:q(t_{0})=q_{0}$ respectively.

Use is frequently also made of the density parameter
$\Omega =\varrho/\varrho_{cr}$
where the critical density is

\be
\label{rhocr}
\varrho_{cr} = \frac{3 H^{2}}{8\pi G}.
\ee

Eq.\ref{freq2} may be written also in the form:

\be
\label{dec}
q = \frac{1}{2}\Omega \left(1 + \frac{3p}{\varrho c^2} \right)
\ee
where $\Omega, p ,\varrho$ are the total quantities (see Eq.\ref{emt}).

The old standard model has the following parameters

\be
\label{old}
\Omega_0 = \Omega_{(m)0} = 1, \;\; \Omega_v =0, \;\; q_0=0.5
\ee

The new version of  SM which is currently accepted  is

\be
\label{new}
\Omega_0 \approx 1 , \;\;\Omega_m \approx 0.2,\;\;
\Omega_v \approx 0.8, \;\; q_0 \approx -0.7, \;\;H_0=65\pm10\;km /s Mpc
\ee

This means that the expansion of the present universe is
accelerated and that the dominant force in the universe is
cosmological antigravity of the vacuum (see discussion by Krauss,1999).
Sandage's problems N.18 and N.20 
are related to the value of parameters $q_0$
and $\Omega_0$, and recent observations of distant supernovae
now specify their values (see discussion in Sec.4.7).

\subsection{Mattig's distance-redshift relation}

In the case of a matter dominated FLRW model there is very important
explicit relation between cosmological redshift and metrical
distance at present epoch $t=t_0$. The relation was firstly derived
by Mattig (1958) and has the form (see e.g. Peacock,1999)

\be
\label{matt}
l_m(z,q_0)=S_0I_k(\chi) =
\frac{c}{H_{0}}\frac{zq_{0}+(q_{0}-1)
((2q_{0}z+1)^{1/2}-1)}{q_{0}^{2}(1+z)}
\ee
where $l_m(z,q_0)$ is the cylindrical metric distance,
$\chi$ is the spherical comoving coordinate distance,
$q_0$ is the deceleration parameter, $z$ is the
cosmological redshift, $I_k = \sin\chi,\chi,\sinh\chi$
for $k= +1,0,-1$ respectively, and
the scale factor is

\be
\label{s0}
S_{0}=\frac{c}{H_{0}}\sqrt{\frac{k}{2q_0 - 1}}
\ee

To calculate the internal metrical distance $r_m$ for a
known Hubble constant, deceleration parameter and redshift,
one must then use also the relation (Eq.\ref{mdrel}) between $l$ and $r$.

\subsection{Observable quantities}

The basic relation for the calculation of different
observable quantities within FLRW model
is the connection between
metric $\:r_{m}$, angular $\:r_{a}$ and luminosity
$\:r_{lum}$ distances in the expanding universe:

\be
\label{mal}
r_{m}=r_{a}(1+z) =\frac{r_{lum}}{(1+z)}
\ee

Using the Eq.\ref{mal} one may calculate such theoretical
predictions as angular size-redshift, magnitude-redshift,
count-magnitude and count-redshift relations.

Classical cosmological tests, such as $\Theta (z),\; m(z),\; N(m)$,
are based on  these relations and actually give practical tools
for estimation of the observed values of the main cosmological parameters
(see e.g. Baryshev et al.,1994; Peacock,1999).

\subsection{Successes of the Standard Model}

According to modern cosmological textbooks
(see e.g. Peebles,1993; Peacock,1999)
the Standard Model is
the homogeneous FLRW model of the universe,
which begins from a singularity
and has expanded in a near homogeneous way from
a denser hotter state when the  cosmic background radiation was
thermalized.

There is definite success in the application
of the SM to the  observed Universe.
Indeed, there are no
gravitational, photometric and thermodynamic
 paradoxes in the SM, because the age
 of the Universe is finite and rather small,
equal to the age of a solar-like normal star.

In the SM, space has been filled  with blackbody radiation, the
cosmic microwave background radiation (CMBR).
The number of CMBR photons per unit volume at redshift $z$ is

\be
\label{48}
n(z)= n_{0}(1+z)^3
\ee
when $\:n_{0}$ is the present value of the number density.
As the universe expands CMBR preserves a black body spectrum with
the temperature

\be
\label{49}
T(z) = T_{0}(1+z)
\ee
where $\:T_{0}$ is the present temperature
of the CMBR. Eq.\ref{48} and Eq.\ref{49} are used
 back to $\:z\sim10^{10}$. The observed
thermal spectrum of the CMBR is considered as the greatest
success of the SM.

In the SM, the Universe was hot and dense enough
to drive thermonuclear reactions that changed the
 chemical composition of the matter. The values
of the abundances left over from this hot epoch
depend on the cosmological parameters.
Knowing the present temperature and
 assuming a value for the present matter
 density, the thermal history
of the Universe is fixed.
If the matter is uniformly
 distributed and lepton numbers are comparable
to the baryon number, this is sufficient to fix the
 final abundances of the light elements.
The observed light element abundances
of $\:^4He,\:^2H,\:^3He$ and $\:^7Li$ are
 in good agreement with SM predictions.

\section{ Conceptual problems of the Standard Model}

In parallel  with the successes of the SM there are several
deep conceptual puzzles which have no convincing explanation yet
and which need more careful analysis at the present time
when foundation of the SM is under consideration.

\subsection{The nature of the  expansion of space}

According to SM the space of our Universe is described by
RW metric (see Eq.\ref{rw1} and Eq.\ref{rw2}).
In mathematical language our 3-dimensional space at the fixed
cosmic time is just a hypersphere in 4-dimensional 
embedding Euclidean space. Hence the space expansion
simply means that the radius of the hypersphere grows with time
and  3-dimensional volume of the space continuously increasing,
i.e. for an internal 3-dimensional observer the space
is continuously created. The puzzling physical problem is that
the space in physics is not empty
but it relates to the pysical vacuum, so 
the physics of space creation needs to be explained.

Another problem is how to measure the space expansion.
Indeed if our Galaxy does not expend then it is hopeless
problem to verify this new physical phenomenon by laboratory
experiments and one has to only  believe in the theoretical
interpretation of cosmological redshift.

\subsection{Recession velocities of distant galaxies more
than velocity of light}

The {\em exact relativistic} expression for recession velocity, or the
"space expansion" velocity,
or the rate of increasing of the metric distance $r$,
for a body with fixed $\:\chi$ directly
follows from Eq.\ref{dist1} :

\be
\label{expvel}
V_{exp} = \frac{dr}{dt} =  \frac{dS}{dt} \chi = \frac{dS}{dt}
\frac{r}{S} = H(t) r = c \frac{r}{r_{H}}
\ee
where $\:H(t)= \dot{S}/S $
is the Hubble constant (actually depends on time)
and $\,r_{H} = c/H(t)$ is the Hubble distance at the time $\,t$.

The exact relativistic velocity - distance  relation is 
Eq.\ref{expvel} and it is linear for all distances $\,r$. It means
that for $\,r>r_{H}$ we get $\,V_{exp}>c$ and the question arises
why general relativity violates special relativity. The usual answer
is that the space expansion velocity is not ordinary velocity of a
body in space, hence it has no ordinary limit by the velocity of
light. This question is tightly connected with  fact  
 mentioned above ,
that space expansion redshift and Doppler redshift are quite
different physical phenomena (see discussion in Harrison, 1993).

\subsection{Cosmological Friedmann force}

Friedmann's equation (Eq.\ref{freq2}) in fact presents
the {\em exact relativistic}
cosmological Friedmann force acting on a test galaxy with mass $m$
placed at a distance $r$ from any fixed point at the origin of
coordinate system:

\be
\label{frforce}
F_{Fr}(r) = m \frac{d^2r}{dt^2} = -\frac{GmM_g(r)}{r^2}
\ee
It looks like the usual Newtonian
equation of motion of a test particle. Such a similarity was
first found by Milne(1934) and McCrea\& Milne(1934) and 
created a problem in cosmology because Eq.\ref{frforce} has no such
relativistic restrictions as limit by velocity of light and
general retarded response effects. The root of the puzzle lies
in the derivation of Friedmann's equation, which utilizes the
comoving coordinates $r$ and synchronous universal cosmic time $t$.

For example, the critical density (Eq.\ref{rhocr})
of the FLRW universe does not depend on the velocity of light
and simply is the Newtonian pulsation formula. The superluminous
expansion velocity (Eq.\ref{expvel}) also is a consequence of this
{\em non-relativistic} character of Friedmann's equation.

\subsection{ Continuous creation of gravitating mass}

The most puzzling property of the FLRW model is the dependence
of gravitating mass in Eq.\ref{freq2} on the cosmic time $t$.
Indeed, in the case of ordinary matter the density
$\varrho_m \sim r^{-3}$ and the gravitating mass Eq.\ref{mgm}
does not depend on time. However in the case of radiation the
density is  $\varrho_r \sim r^{-4}$ and the gravitating mass of
radiation will be

\be
\label{mgr1}
M_r(r) = \frac{4 \pi }{3} 2 \varrho_r r^3 \;\sim r^{-1}(t)
\ee

This means that mass of radiation continuously disappeared in
the expanding universe. As it is noted by Peebles(1993, p.139):
"The resolution of this apparent paradox is that ... there is
not a general global energy conservation law in general
relativity theory."

The next strange example is the vacuum, where the density
$\varrho_v$ is a constant in time, so the gravitating mass of the
vacuum will be

\be
\label{mgv1}
M_v(r) = - \frac{4 \pi}{3} 2\varrho_v r^3 \;\sim r^3(t)
\ee

This means that vacuum antigravity continuously increase in time
due to continuous creation of gravitating (actually
"antigravitating") vacuum mass.

\subsection{ Cosmological pressure is not able to produce a work}

It was noted by Harrison(1981; 1995) that in a homogeneous
unbounded expanding  universe there is no pressure gradient and so
the first law of laboratory thermodynamics

\be
\label{1law}
dE/dt + p\;dV/dt\;=0
\ee
is not applicable. Indeed in the case of the FLRW model we may
imagine the whole universe partitioned into macroscopic cells,
each of comoving volume $V$, and all having contents in identical
states. The $-p\,dV$ energy lost from any one cell cannot reappear
in neighboring cells because all cells experience identical losses.
So the usual idea of an expanding cell performing work on its
surroundings cannot apply in this case. As Edward Harrison
emphasized:  "The conclusion, whether we like it or not, is obvious:
energy in the universe is not conserved" (Harrison, 1981, p.276).

\subsection{Cosmological gravitational frequency shift}

In 1947 in the classic paper "Spherical symmetrical models in
general relativity" by Sir Hermann Bondi it was shown that,
at least for small redshifts,
the total cosmological redshift of a distant body is due to
two causes: the velocity shift (Doppler effect) due to the relative
motion of source and observer, and the global gravitational
shift (Einstein effect) due to the difference between the potential
energy per unit mass at the source and at the observer.

It means  that the spectral shift   depends not only on the conditions
at the source and at the observer but also on the distribution of
matter in the intervening space around the source. 
In the case of small distances
Bondi derived simple formula for redshift which is simply
the sum of Doppler and gravitation effects, and which explicitly
showed that "the sign of the velocity shift depends on the sign
of $v$, but the Einstein shift is easily seen to be towards the red"
(Bondi,1947,p.421).

Hence according to Bondi the cosmological gravitational frequency
shift is redshift. It was shown by Baryshev et al.(1994) that from
Mattig's relation (Eq.\ref{matt}) it follows directly for the case
of $z<<1, \,v/c \approx x = r/r_H$ that

\be
\label{zcos}
z_{cos}\approx x + \frac{1+q_0}{2}x^2 =
(\frac{v}{c} + \frac{1}{2}\frac{v^2}{c^2}) +
\frac{q_0}{2}x^2
\ee
is the sum of Doppler and gravitational redshifts:

\be
\label{zcos1}
z_{cos} \approx z_{Dop} + z_{grav}
\ee
where the cosmological gravitational redshift is

\be
\label{zgrav}
z_{grav} = \frac{\Delta\varphi(r)}{c^2} =
\frac{1}{2}\frac{GM(r)}{c^2 r} = \frac{1}{4}\Omega_0 x^2
\ee

Here $r$ is the distance between the observer and the source,
and the source is in the center of the sphere.

An ambiguity arises when one consider the observer at the center
and a galaxy at the edge of the sphere. In this case one may conclude
that cosmological gravitational shift is blueshift (see Zeldovich\&\,
Novikov,1984, p.97 and Peacock,1999, problem 3.4).

Interesting that for a fractal matter distribution 
in wich $M(r) \sim r^D$ with fractal
dimension $D=2$ the cosmological gravitational redshift gives
the linear distance-redshift relation and becomes an observable
physical phenomenon.

\subsection{Friedmann-Holtsmark paradox}

According to  Friedmann's equation  there is
the cosmological force Eq.(\ref{frforce}) acting on a galaxy situated 
at the distance $r$ from another fixed galaxy. 
The value of the cosmological force
is equal to the value of Newtonian force for the finite spherical
ball with radius $r$ around the fixed galaxy. So this cosmological
force increases up to infinity when a galaxy is infinitely far.
Moreover the Friedmann force determines the dependence on time
of the scale factor $S(t)$, so it plays a fundamental role in the SM.

This is in apparent contradiction with the well known Holtsmark
result for the probability density of the force acting between
particles in infinite Euclidean space in the case of $1/r^2$
behavior of the elementary force (see Holtsmark,1919;
Chandrasekhar,1941).
Due to isotropy of the particle
distribution the average force is equal to zero and there is
the finite value of the fluctuating force, which is determined by
the nearest neighbor particles. Hence in infinite Euclidean space
with homogeneous Poisson distribution and Newtonian gravity force
there is no global expansion or contraction, but there is
the density and velocity fluctuations caused by gravity force fluctuations.

Recently it was found by de Vega\&Sanchez(1999) that the ground state
of the self-gravitating Newtonian gas is the fractal mass distribution with
fractal dimension $D\approx 2$.
Probably the final state of initially Poissonian self-gravitating gas 
will be this deVS ground state. Future N-body simulations can check
this possibility.

\subsection{The problem of the cosmological constant}

The claim "New observations have smashed the old view of our universe"-
 opened the January 1999 issue of Scientific American,
devoted to special report on revolution in cosmology because of
new observations of very distant supernovae. Two independent
groups of astronomers (Riess et al.,1998; Perlmutter et al.,1999)
have constructed the magnitude-redshift
relation for about fifty SNIa in distant galaxies within redshift
interval $0.1 - 1.0$. The result was completely unexpected because
it showed significant deflection from prediction of the standard
model for $\Omega_m = 1$. To fit observational data one needs
positive cosmological constant giving $\Omega_v \approx 0.8$.
So what Einstein called "the biggest blunder of my life" now
became the biggest news in cosmology.

This is a quite unexpected solution of  Sandage's 18th problem
because in the framework of FLRW model it means that the observed 
universe is accelerating under a mysterious repulsive force which 
dominates the dynamics of the universe.
Within the old version of SM the cosmological constant "naturally"
had zero value and so did not participate in present time dynamics
of the universe.
 
Using Eqs.\ref{emt}, and \ref{rhocr}
we get for the observed vacuum density

\be
\label{rhov1}
\varrho_v = 6.2\cdot10^{-30}\Omega_v h^2_{60}(\frac{g}{cm^3})
\ee
where $h_{60}=H_0/60(km/s\; Mpc)$ is normalized Hubble constant.

This result is very hard to explain theoretically. Indeed a naive
theoretical estimation of the energy density of the vacuum includes
the sum of zero point energies of all physical quantum fields, which
must be calculated up to certain high energy cutoff $\,k_{max}$

\be
\label{lambda5}
\varrho_v \approx\frac{\hbar}{c} \frac{k_{max}^{4}}{16 \pi^{2}}
\ee

If one takes as the cutoff the Planck energy $E_{Pl}=m_{Pl} c^2$
with $m_{Pl} = \sqrt{\frac{\hbar c}{G}}$ then
$\,k_{max} \approx E_{Pl}/\hbar c$ and the theoretical value of the
vacuum density will be

\be
\label{rhov2}
\varrho_v \approx \varrho_{Pl} = \frac{c^5}{G^2\hbar}
= 5.46\cdot 10^{93} \frac{g}{cm^3}
\ee

Hence theoretical expectation for the cosmological
constant exceeds the observed value by  $\,123$ orders of
magnitude.

Weinberg(1989) considered various possible solutions of this
problem based on different approaches:  all these approaches show
that the cosmological constant problem have great impact on other
areas of physics and astronomy. Weinberg note: "More discouraging
than any theorem is the fact that many theorists have tried to invent
adjustment mechanisms to cancel the cosmological constant, but
without any success so far".

Another problem connected with non-zero cosmological constant or
cosmological scalar field  
({\it "quintessence"}) was mentioned above puzzle of the
continuous creation of the corresponding gravitating mass in
Friedmann's equation (Eq.\ref{freq2}). Indeed, the density of the vacuum
does not change with time, hence its mass within comoving radius $r(t)$
grows with time as $r^3(t)$. In the case of quintessence the dipendence
on time is defined by production $\varrho_v(t)r^3(t)$ as it follows
from Eq.\ref{mgv1}.

At last, the observed  approximate equality 
of matter and vacuum densities at present epoch
leads to a puzzling "fine tuning" or
coincidence: the density of ordinary matter rapidly decreases
as the universe expands but the density of vacuum is fixed,
so why, despite these opposite behaviors, do the two densities
have nearly the same value today?

\section{Cosmological Principle}

One of the fundamental elements of modern cosmology is the
Cosmological Principle (CP) and it is very important to understand
its different formulations and applications. Sometimes misleading
claims appear in the literature about the CP,  especially when a
fractal matter distribution is discussed.

\subsection{Einstein's Cosmological Principle}

In the section devoted to the SM we already mentioned a formulation of
Einstein's CP, which states that the universe is homogeneous
(constant density) and isotropic (the same in all directions).
In modern cosmological textbooks there is also another more weak
formulation of the CP: the universe has no center and is isotropic in
any place, or that humans are not privileged observers.

There is a widely spread opinion that from isotropy and the absence of
a prefered centre
one may deduce homogeneity of the universe (see e.g. Peacock,1999,
p.65). Strictly speaking this inference is true only for continuous matter
distribution and not true for discrete sets (e.g. fractals).

\subsection{Fractality of observed galaxy distribution}

For a long time, astronomers used only photographic plates
of the sky as the basic means for the galaxy structures
studies without no direct observations of the
3-dimensional large-scale matter distribution.

Recently, several 3-dimensional maps of galaxy
distribution  have become available, based on massive
redshift measurements.
Surveys such as CfA, SSRS, Perseus-Pisces, IRAS,
LEDA, APM-Stromlo, Las Campanas, and ESP for galaxies, and Abell and
ACO for galaxy clusters have detected remarkable
structures such as filaments, sheets and voids.
The galaxy maps now
probe scales up to $200h_{60}$ Mpc and they show that the
large-scale structures are common features of the local universe.

Pietronero and collaborators  (see Pietronero,1987 and
review by Sylos Labini et al.,1998
for a comprehensive discussion of the subject), by using the methods
of modern statistical physics, have shown that, in the various surveys,
galaxy distribution exhibits fractal behavior with dimension
$D\approx 2$ at least up to $200 Mpc$
and the size of the upper cutoff, if it exists, must
be more than $200 Mpc$ (see web page devoted to debate on fractality
of galaxy distribution http://pil.phys.uniroma1.it ).

It is important to note that according to recent $N(r)$ count-distance
analysis of the complete sample of KLUN spiral galaxies
by Teerikorpi et al.(1998), it was shown that number of galaxies
increases as $r^{(2.2\pm 0.2)}$ up to the distance $r\approx 200$ Mpc.
This result  solves the old controversy between the observed local
inhomogeneous galaxy distribution and $N(m)$ count-magnitude relation
with $0.6m$-law. Now direct $N(r)$ count-distance relation is in 
accordance with a fractal galaxy distribution up to 200 Mpc.

\subsection{Mandelbrot's Cosmological Principle}

Homogeneity of visible matter
up to a hundred Mpc is disproved now by direct observations
of the spatial galaxy distribution. But is the Cosmological Principle
true?   From the Einstein's CP of homogeneity and isotropy  it
follows that the universe is the same in every place and in every
direction. However it is possible to formulate a more general CP
which possess these properties in an inhomogeneous discrete matter
distribution.

This is the Mandelbrot's Cosmological Principle, which states that
in a statistical sense an inhomogeneous fractal matter distribution
in the universe is isotropic around any structure point and has no
center (Mandelbrot,1977; 1982). In the fractal universe density of
matter depends on the scale of statistical averaging and may be even
zero for infinite distances. So the fractal universe is not
"An unprincipled Universe" as was claimed by Coles(1998), but
is simply a universe obeing a more general cosmological principle.

Isotropy of a fractal distribution means that usual arguments for
homogeneity based on observed isotropy are not generally valid.
The only convincing test of fractality is the direct study of
space galaxy distribution by measuring redshifts for huge number
of galaxies. Such projects as 2dF and Sloan will show soon the true
nature of visible matter distribution up to the
scales approaching the Hubble radius.

\subsection{Einstein-Mandelbrot's Cosmological Principle}

There is some astrophysical evidence for possible homogeneity
at very large scales close to the Hubble radius. For example from
the isotropy of CMBR it follows that at scales about several
thousands Mpc electromagnetic radiation fills the universe
homogeneously, because of photons can not cluster as usual matter.

If the fractal distribution of ordinary matter extends up to the
scales where density of radiation dominates then one has the
universe which is essentially fractal inside the Hubble radius and
which is essentially homogeneous outside the Hubble radius.
For the case one may say that the Einstein-Mandelbrot's Cosmological
Principle of no center and statistical isotropy valid at all scales.

\section{Sandage's 21st problem: Hubble - de Vaucouleurs paradox}

Discovery of a fractal galaxy distribution within the scales of about
200 Mpc has created a new puzzle in cosmology.
Indeed, the SM assumption of homogeneity
"leads to the prediction of Hubble's law - that the apparent
recession velocity of a galaxy is proportional to its distance - for
that is the only expansion law allowed by homogeneity" (Peebles,
1993, p.5).
Consequently, without direct information about real spatial
distribution of matter in the universe, it was usually claimed that
from linear Hubble law it follows that the universe is homogeneous
just from the scales where the linearity of the Hubble law was found.

In an important earlier paper, Sandage et al.(1972) were the first
to note the surprising co-existance of the linear Hubble law and
local inhomogeneites. Actually they used the observed linearity
of Hubble law at small distances as a strong argument against
de Vaucouleurs' hierarchical universe. Later in 1995 in the list
of "Astronomical Problems for the Next Three Decades" Sandage
devoted the 21st problem to this subject in the form of the question:
"Are there significant velocity deviations from the pure cosmological
expansion?".

\subsection{Statement of the HdeV paradox}

According to modern observations
based on Cepheid distances to local galaxies, Tully-Fisher
distances from the KLUN program, and Supernovae Ia distances
(see Teerikorpi,1997; Ekholm
et al.,1999) the linear Hubble law is well established starting from
scales of about 1 Mpc.

But, as we have already mentioned,
studies of the 3-dimensional galaxy universe have shown
that de Vaucouleurs' prescient view on the matter distribution
(de Vaucouleurs,1970)
is valid at least in the range of scales  $\sim 1 \div 200 \;Mpc$
(Sylos Labini et al.,1998).

The Hubble and de Vaucouleurs laws describe very different
aspects of the Universe, but both have in common universality and
observer independence.
This makes them fundamental cosmological laws and it is important
to investigate the consequences of their coexistence at
the same length-scales (see Baryshev et al.,1998).

A puzzling conclusion is that the strictly linear
redshift-distance relation is observed deep inside the
fractal structure, i.e. for distances less than the homogeneity
scale $r_{hom}$:
\be
\label{hdev}
(\;r \;<\;r_{hom}\;)\;\;\& \;\;(\;cz=H_0r\;)
\ee

This empirical fact presents a profound
challenge to the standard model in which
homogeneity is the basic explanation of the Hubble law, and "the
connection between homogeneity and Hubble's law was the first success of
the expanding world model" (Peebles et al.,1991).

In fact, within the SM one would not expect any neat relation of
proportionality between velocity and distance
for nearby galaxies, which are members of large scale structures.
However, contrary to the expectation, modern data show a good linear
Hubble law even for nearby galaxies. It leads to a new observationaly
established puzzling fact that the linear Hubble law is not
a consequence of a homogeneity of visible matter, just because
the visible matter is distributed inhomogeneously.

\subsection{Possible solutions of the HdeV paradox}

Up to now several possible solutions of the HdeV paradox
have been suggested. The first one
(Baryshev et al. 1998; Durrer\&Sylos Labini,1998) is based on the
assumption of the existence of uniformly distributed dark matter
starting just from the halos of galaxies, in this case the standard
FLRW solution exists. However, then the fractal distribution of
luminous matter (galaxies) can appear only from a special choice of
initial conditions and hence has no fundamental meaning.

The second solution is to accept a very low value for the
global average density (Baryshev et al. 1998; Humphreys et al. 1998;
Gromov et al.,1999).
However in this case when the value of the upper
cut off scale of the fractal structure is large, the low density
contradicts the available estimates of the density of the
baryonic luminous and dark matter.

Other solutions  of the HdeV paradox are  based on
the more general than FLRW cosmological models.
For instance Lemaitre-Tolman-Bondi (LTB) models are
exact nonlinear solutions of Einstein's equations
under the assumptions of spherical symmetry, pressureless
matter and no spherical layers intersecting.
In the frame of the LTB cosmological models
nonsimultaneous bang time  (Gromov et al.,1999) and
$\Lambda$-term (Baryshev et  al.,1999)
allow the  linear Hubble law to be compatible with a fractal
structure having an upper cut off.

A very different possibility to solve the HdeV paradox comes
from recent discovery  by de Vega \& Sanchez (1999), that
self-gravitating (via Newtonian gravity) N-body systems have a
quasi-equilibrium state which is fractal in its structure with a
fractal dimension of about 2. So, self-gravity naturally leads to
fractality and the actual problem is how to explain the appearance of the
Hubble law inside this structure. As it was shown by Baryshev,1981
(see also Baryshev et al.,1994; 1998) the cosmological gravitational
redshift effect gives the linear redshift-distance relation just
for fractal structure with $D=2$, which is actually observed
at least up to scales about 200 Mpc.
For such a model the main problem is a high value of dark matter
coupled with fractal visible matter needed for explanation of the
observed value of the Hubble constant.

\section{Quantum nature of  gravity force}

The roots of many of the  conceptual problems of modern cosmology
discussed above actually lie in the  gravity theory. In fact, 
all fundamental forces in physics (strong, weak, electromagnetic)
are quantum in nature,
(i.e. there are quanta of corresponding fields which carry
the energy-momentum of physical interactions),
while GR presents the geometrical interpretation of gravity force
(i.e. the curvature of space itself but not a matter in space) 
which, as it is well known,
exclude the concept of localizable gravity energy. This is why the main
problem of GR is the absence of the energy of the gravity field
or pseudo-tensor character of gravity EMT (see Landau\& Lifshitz,1971; 
and for recent attempt
to construct gravity EMT see Babak\&Grishchuk,1999). Together
with GR the energy problem comes to cosmology and is the cause of
some conceptual problems of SM.

The quantum field approach to gravity force was considered by
Feynman(1971) in his "Lectures on Gravitation. 
Within the field approach, the gravity is a kind of matter, i.e. the
tensor field in Minkowski space and 
this means that its quanta - gravitons carry the energy-momentum 
of the gravitational interaction. As Feynman emphasized
"the geometrical interpretation is not really necessary or
essential to physics" (Lecture 8, p.110) and the Field Gravity
Theory (FGT) may be constructed with usual field-theoretical
technique. This means that Minkowski space allows to define
EMT of gravity field and conservation laws without "pseudo"-problems,
and also utilize usual quantum mechanics and quantum field theory.
The main advantage of the field gravity theory is that it gives
positive and localizable energy density of gravitational field
which allows to get gravitons as the energy quanta of the field.

In the case of the weak field approximation
both theories give the same predictions for classical relativistic
gravity effects. But in the scope of FGT there are also new
relativistic effects even in weak fields and profundly different
predictions in the case of strong gravity fields.
For instance, it can be shown, that within FGT the positive
energy density of gravity field exclude the possibility of
black holes and in cosmology there is an expansion of matter
in space but there is no expansion of space. So observed 
cosmological redshift may be related to the Doppler effect
and the cosmological gravitational redshift.
A general discussion of the geometrical and field approaches to gravity
may be found in  Baryshev(1996).

The modern state of gravity theory and experiments was analyzed by Damour(1999),
who emphasized that existing now tests of GR does not exclude
a more general quantum gravity theory which may have very different
predictions for strong field effects. In particular, possible
existence and observational tests for scalar gravitational field
has been discussed by Damour(1999) and Baryshev (1995;1996;1997).
The problem of non-zero mass for the graviton was analyzed by
Visser(1998), who showed that in this case the strong field effects
and cosmological solutions will differs dramatically from GR.

It is important to note that study of the scalar part 
of the gravitational field and the mass of the graviton is not
an "academic" problem but has practical importance,
because of the corresponding theories wil be experimentally tested
in near future by using the gravitational wave observatories
(such as LIGO and VIRGO) which start to operate in two years.

\section{Conclusions}

Two major building blocks of modern cosmological models are
the Cosmological Principle and the  Theory of Gravitation.
Correspondingly the main conceptual problems of cosmology are related to
studies of large scale matter distribution and
physics of the gravitational interaction.
There are fundamental problems in cosmology which are still unsolved
and even have not yet been analyzed, so the opinion that "cosmology is
solved" is a dream far from reality. Moreover deep conceptual puzzles
which we have discussed above actually leave no room for "cosmologists'
arrogance" (see Turner, 1999b) with existing standard model.
Main conclusions of this report are following:

\begin{itemize}

\item The time for Fractal Cosmology is coming, so the old
Cosmological Principle of Homogeneity must be replaced by
the new more general Cosmological Principle of Fractality.
The new Cosmological Principle  is fully compatible with the
reasonable requirements of the  equivalence of all the observers
and the condition of local isotropy around any structure point.
The case of Einstein-Mandelbrot universe where essential fractal
matter distribution at small and intermediate scales becomes
homogeneous at very large scales is a particular model
of Fractal Cosmology.

\item The paradox of linear Hubble law within the fractal
visible matter distribution implies the high value of
homogeneous dark matter, or very low value of asymptotic
FLRW background, or application of more general cosmological
models such as Lemaitre-Tolman-Bondi model.

\item  Modern relativistic quantum field theory shows that
future gravity theory will be more general than general relativity.
Within the framework of quantum field approach to gravity
there are such unexplored possibilities as the scalar part
of the gravity field and nonzero rest mass of the graviton.

\item  Crucial future
observational tests are needed to make distinction between
rival cosmological models. Among them : fractal dimension and maximum
scale of fractality of spatial galaxy distribution (2dF, SLOAN); detection
of gravitational waves (LIGO, VIRGO); physical properties of high
redshift galaxies, radio galaxies and quasars (HDFS).

\end{itemize}

{\bf REFERENCES}\\

Babak S., Grishchuk L., 1999, gr-qc/9907027

Baryshev Yu.V., 1981, Izvestiya SAO 14, 24

Baryshev Yu., 1995, in "First Amaldi Conference on Gravitational Wave
Experiments" p.251 (World Scientific); (gr-qc/9911081)

Baryshev Yu., 1996, Gravitation 2, 69; (gr-qc/9912003)

Baryshev Yu., 1997, Astrophysics 40, N3, 244

Baryshev Yu., Gromov A., Teerikorpi P., 1999 (in preparation)

Baryshev Yu.V \& Teerikorpi P. Discovery of Fractal Universe:
paradigms upheaval in cosmology  (in preparation)

Baryshev, Yu., Sylos Labini F., Montuori, M., Pietronero, L., 1994,
Vistas in Astronomy 38, 419; (astro-ph/9503074)

Baryshev, Yu., Sylos Labini F., Montuori, M., Pietronero, L., Teerikorpi, P.,
1998, Fractals 6 No.3, 231; (astro-ph/9803142)

Bondi, H., 1947, MNRAS 107, 410

Chandrasekhar, S., 1941, Astrophys.J. 94, 511

Coles P., 1998,  Nature 391, 120

Damour T., 1999, gr-qc/9904057

de Vaucouleurs, G., 1970, Science 167, 1203

de Vega, H., Sanchez, N., 1999, hep-th/9903236

Durrer R., Sylos labini F., 1998, A\&A Lett., 339, L85

Ekholm T., Lanoix P., Paturel G., Teerikorpi P.,
1999, A{\&}A (in press)

Feynman R., 1971, Lectures on Gravitation (California Institute of Technology)

Gromov A., Baryshev Yu., Suson D., Teerikorpi P.,
1999, A{\&}A (submitted); (gr-qc/9906041)

Humphreys, N.P., Maartens, R., Matravers, D., 1998,
Class. Quantum Grav 15, 3041

Harrison E., 1981, Cosmology ( Cambridge University Press)

Harrison E., 1993, Astrophys.J., 403, 28

Harrison E., 1995, Astrophys.J., 446, 63

Holtsmark, J., 1919, Ann. d. Phys. 58, 577

Krauss L., 1999, Sci.Am., 280, 34

Landau, L.D., Lifshitz, E.M., 1971, The Classical Theory of Fields
(Pergamon Press)

Mandelbrot B., 1977, Fractals:Form, Chance and Dimension
(W.H.Freeman)

Mandelbrot B., 1982, The Fractal Geometry of Nature (W.H.Freeman)

McCrea W., Milne E., 1934, Quart.J.Mathem., 5, 73

Milne E., 1934, Quart.J.Mathem., 5, 64

Peebles P.J.E., 1993, Principles of Physical Cosmology
(Princeton Univ.Press)

Peebles P.J.E., 1999, Publ.A.S.P., 111, 274

Peebles P.J.E. et al.,1991, Nature, 352, 769

Perlmutter S. et al., 1999, ApJ. 517, 565

Pietronero L., 1987, Physica, A144, 257

Riess A. et al., 1998, A.J. 116, 1009

Sandage A., 1995, Proc. of the conf."Key Problems in Astronomy and Astrophysics"

Sandage A., Tammann G., Hardy H., 1972, Ap.J., 172, 253

Sylos Labini F., Montuori, M., Pietronero, L.,
1998, Phys.Rep. 293, 61

Teerikorpi P., 1997, Ann.Rev.Astron.Astrophys. 35, 101

Teerrikorpi P. et al., 1998, A\&A, 334, 395

Turner M., 1999a, Publ.A.S.P. 111, 264

Turner M., 1999b, Reflection 2000 (the University of Chicago site)

Visser M., 1998, gr-qc/9705051

Weinberg S., 1972, Gravitation and Cosmology (John Wiley \& Sons)

Zeldovich Yu.B., Novikov I.D., 1984, Relativistic Astrophysics Vol.2
(The University of Chicago Press)

\end{document}